\documentclass[twocolumn,aps,showpacs,pra,10pt,superscriptaddress]{revtex4-1}
\usepackage{amsmath}
\usepackage{amssymb}
\usepackage{graphicx}
\usepackage{hyperref}
\usepackage{latexsym}
\usepackage{color}

\begin{document}
\title{Violation of Bell's inequality for phase singular beams}
\author{Shashi Prabhakar} \email{shaship@prl.res.in}
\affiliation{Physical Research Laboratory, Navrangpura, Ahmedabad, India. PIN 380009.}
\affiliation{IIT Gandhinagar, Chandkheda, Ahmedabad, India. PIN 382424.}
\author{Salla Gangi Reddy} 
\affiliation{Physical Research Laboratory, Navrangpura, Ahmedabad, India. PIN 380009.}
\author{A Aadhi} 
\affiliation{Physical Research Laboratory, Navrangpura, Ahmedabad, India. PIN 380009.}
\affiliation{IIT Gandhinagar, Chandkheda, Ahmedabad, India. PIN 382424.}
\author{Chithrabhanu Perumangatt} 
\affiliation{Physical Research Laboratory, Navrangpura, Ahmedabad, India. PIN 380009.}
\author{G. K. Samanta} 
\affiliation{Physical Research Laboratory, Navrangpura, Ahmedabad, India. PIN 380009.}
\author{R. P. Singh} 
\affiliation{Physical Research Laboratory, Navrangpura, Ahmedabad, India. PIN 380009.}

\date{\today}

\begin{abstract}
We have considered optical beams with phase singularity and experimentally verified that these beams, although being classical, have properties of two mode entanglement in quantum states. We have observed the violation of Bell's inequality for continuous variables using the Wigner distribution function (WDF) proposed by Chowdhury et al. [Phys. Rev. A \textbf{88}, 013830 (2013)]. Our experiment establishes a new form of Bell's inequality in terms of the WDF which can be used for classical as well as quantum systems.
\end{abstract}


\maketitle

\section{Introduction}
Optical vortices, phase singularities of the electromagnetic field, show interesting classical and quantum properties with a variety of applications \cite{gibson_free-space_2004, molina-terriza_twisted_2007, thide_utilization_2007, bhattacharya_entanglement_2008, bozinovic_terabit-scale_2013}. They are observed as dark spots in bright background. For the vortex of topological charge or order $n$, the azimuthal phase varies as $2\pi n$ in a full rotation around the dark spot \cite{allen_optical_2003}. The sense of rotation of phase provides the sign of its charge \cite{vaity_measuring_2013}. The topological charge $n$ can be considered as an important parameter for such beams. One of the main characteristics of these beams is that they carry an orbital angular momentum (OAM) of $n\hbar$ per photon \cite{allen_orbital_1992}. This particular property has successfully been utilized for particle manipulation \cite{grier_revolution_2003} and quantum information \cite{mair_entanglement_2001, molina-terriza_twisted_2007, fickler_quantum_2012}. Vortex beams have been experimentally realized not only in optics but also with electron beams \cite{verbeeck_production_2010, mcmorran_electron_2011}. These beams form an infinite dimensional basis for applications such as quantum computation and cryptography \cite{mair_entanglement_2001}. Moreover, increase in information entropy with the order of vortices can be utilised to encode more amount of information in these structures \cite{agarwal_spatial_2002, kumar_information_2011}. Here, we present a new aspect i.e. inseparability of position and momentum akin to quantum entanglement, for these fascinating optical structures.

Non-quantum entanglement has been discussed earlier also \cite{spreeuw_classical_1998, lee_entanglement_2004, luis_coherence_2009, simon_nonquantum_2010, chen_single-photon_2010, borges_bell-like_2010, qian_entanglement_2011, ghose_entanglement_2014, toeppel_classical_2014, aiello_quantum--like_2015}, however, there has not been a rigorous experiment showing the violation of Bell's inequality involving inseparability of continuous variables for a classical system. Here, we present our experiment using the theoretical results \cite{chowdhury_nonlocal_2013} to show that inseparable position and momentum variables of a phase singular beam do violate Bell's inequality. A brief theoretical discussion on the Bell's inequality violation for optical vortices has been given in Section \ref{sec:theory}, experiments performed in Section \ref{sec:experiment} and results in Section \ref{sec:result}. Finally we conclude in Section \ref{sec:conclusion}.

\section{Theoretical Background}\label{sec:theory}
The study of WDF for classical beams has been found to be very useful since it can provide coherence information in terms of the joint position and momentum (phase-space) distribution for a particular optical field \cite{tombesi_all-optical_1996, pratap_singh_wigner_2006}. The electric field of a optical vortex of order $n$ and centered at the origin can be written in terms of Laguerre Gaussian (LG) modes
\begin{eqnarray}\label{eq:LG}
    E_{nm}(r,\phi,z)&=&\frac{C^{LG}_{nm}}{w(z)} \left( \frac{r \sqrt{2}}{w(z)} \right) ^{|n|} \exp \left(-\frac{r^2}{w^2(z)} \right) \nonumber \\
    &\times &L_m^{|n|} \left(\frac{2r^2}{w^2(z)}\right) \exp\left( i k \frac{r^2}{2 R(z)}\right)\exp(i n \phi) \nonumber \\
    &\times &\exp\left[i(2m+|n|+1)\zeta(z)\right],
\end{eqnarray}
where $L_m^{|n|}$ is the generalized Laguerre polynomial with $m$ and $n$ as radial and azimuthal indices. $C^{LG}_{nm}$ is the normalization constant, $w(z)$, $R(z)$ and $\zeta(z)$ are beam parameters and $r$, $\phi$ are radial and azimuthal coordinates respectively. For $n\neq0$, such beams contain an azimuthal phase dependence of $\exp(in\phi)$ and a singularity at the center.

The WDF for optical vortex beams can be written as \cite{tombesi_all-optical_1996, simon_wigner_2000}
\begin{eqnarray}\label{eq:wdf}
    W_{nm}(X,P_X&;&Y,P_Y)=\frac{(-1)^{n+m}}{\pi^2} L_n[4(Q_0+Q_2)] \nonumber \\
    &~&\times L_m[4(Q_0-Q_2)] \exp (-4Q_0),
\end{eqnarray}
where \{$X,P_X$\} and \{$Y,P_Y$\} are conjugate pairs of dimensionless quadratures while $Q_0$ and $Q_2$ are
\begin{eqnarray}
    Q_0&=&\frac{1}{4} \left[ X^2+P_X^2+Y^2+P_Y^2 \right], \nonumber \\
    Q_2&=&\frac{XP_Y-YP_X}{2}.
\end{eqnarray}
The scaled variables $X$, $P_X$, $Y$ and $P_Y$ can be defined as
\begin{eqnarray} \label{eq:scaled_shift}
    x(y)&\rightarrow&\frac{w}{\sqrt{2}}X(Y), \nonumber \\
    p_x(p_y)&\rightarrow&\frac{\sqrt{2}\lambdabar}{w}P_X(P_Y)
\end{eqnarray}
and follow $[\widehat{X}, \widehat{P}_X]=[\widehat{Y},\widehat{P}_Y]=i$.

The WDF defined in Eq. \ref{eq:wdf} can be obtained by taking the Fourier transform (FT) of two-point correlation function (TPCF) that is defined as
\begin{eqnarray}
\Phi(x,\epsilon_x;y,\epsilon_y) =&~& \left< E(\epsilon_x+x/2,\epsilon_y+y/2) \right. \nonumber \\
                   &\times & \left. E^\ast(\epsilon_x-x/2,\epsilon_y-y/2)\right>,
\end{eqnarray}
In fact, to determine the WDF in experiment, one measures TPCF only. We produce different orders of vortex beams (LG modes, Eq. \ref{eq:LG} with $m$=0) using spiral phase plate (SPP) \cite{beijersbergen_helical-wavefront_1994} and obtain two-point correlation function using interference between vortices of the same order in a shearing Sagnac interferometer (SSI) \cite{iaconis_direct_1996, cheng_variable_2000}.

For discrete entangled systems, the Bell-CHSH inequality can be written as \cite{bell_einstein-podolsky-rosen_1964, clauser_proposed_1969}
\begin{equation}
B=|S(a,b)+S(a,b')+S(a',b)-S(a',b')| < 2,
\end{equation}
where ($a$, $b$), ($a'$, $b'$) are two analyser settings and $S$($a$, $b$) is the joint probability corresponding to settings ($a$, $b$). The entanglement in quantum systems with continuous variables is characterized by probabilities. For continuous variable systems, the WDF is expressed as the expectation value of a product of displaced parity operators. Banaszek and Wodkiewicz \cite{banaszek_direct_1996, banaszek_nonlocality_1998} have argued that the WDF can be used to derive the analog of Bell's inequality in continuous variable systems.

Considering the transformation $\Pi(X,P_X;Y,P_Y)=\pi^2W(X,P_X;Y,P_Y)$ in dimensionless quadratures, the Bell-CHSH inequality $B$ with chosen points \{$a$,$b$\} $\equiv$ \{$X1,P_{X1};Y1,P_{Y1}$\} and \{$a'$,$b'$\} $\equiv$ \{${X2,P_{X2};Y2,P_{Y2}}$\} can be written as
\begin{eqnarray}\label{eq:eightdimint}
    B&=&\Pi_{nm}(X1,P_{X1};Y1,P_{Y1}) + \Pi_{nm}(X1,P_{X1};Y2,P_{Y2}) \nonumber\\
    &~&+\Pi_{nm}(X2,P_{X2};Y1,P_{Y1})  \nonumber \\
    &~&-\Pi_{nm}(X2,P_{X2};Y2,P_{Y2}) < 2.
\end{eqnarray}

From Eq. \ref{eq:wdf}, the WDF of an optical vortex beam with topological charge $n=$1 ($m$=0 in the present case) can be obtained as
\begin{eqnarray}
    W_{10}(X,P_X;Y,P_Y)=e^{-X^2-P_X^2-Y^2-P_Y^2}\times  \nonumber \\
    \frac{(P_X-Y)^2+(P_Y+X)^2-1}{\pi^2}.
\end{eqnarray}
Choosing $X1=0, P_{X1}=0, X2=X, P_{X2}=0, Y1=0, P_{Y1}=0, Y2=0, P_{Y2}=P_Y$, the Bell-CHSH parameter can be written as
\begin{eqnarray}\label{eq:order1}
    B&=&\Pi_{10}(0,0;0,0)+\Pi_{10}(X,0;0,0) \nonumber \\ 
     &~&+\Pi_{10}(0,0;0,P_Y)-\Pi_{10}(X,0;0,P_Y)
\end{eqnarray}
\begin{eqnarray}
    &=&e^{-P_Y^2}(P_Y^2-1) + e^{-X^2}(X^2-1) - \nonumber \\
    &~&e^{-P_Y^2-X^2}[(P_Y+X)^2-1]-1.
\end{eqnarray}
The maximum Bell's violation considering only two variables $X$ and $P_Y$ is $|B_{max}|\sim$2.17 which occurs at $X\sim$0.45 and $P_Y\sim$0.45. Considering all eight variables from Eq. \ref{eq:eightdimint}, the maximum Bell's violation is $|B_{max}|\sim$2.24 at $X1\sim-$0.07, $P_{X1}\sim$0.05, $X2\sim$0.4, $P_{X2}\sim-$0.26, $Y1\sim-$0.05, $P_{Y1}\sim-$0.07, $Y2\sim$0.26 and $P_{Y2}\sim$0.4. The violation in Bell's inequality can be considered as the non-separability of electric field of the beam used.

\section{Experimental Setup}\label{sec:experiment}
\begin{figure}[b]
\begin{center}
  \includegraphics[width=3.3in]{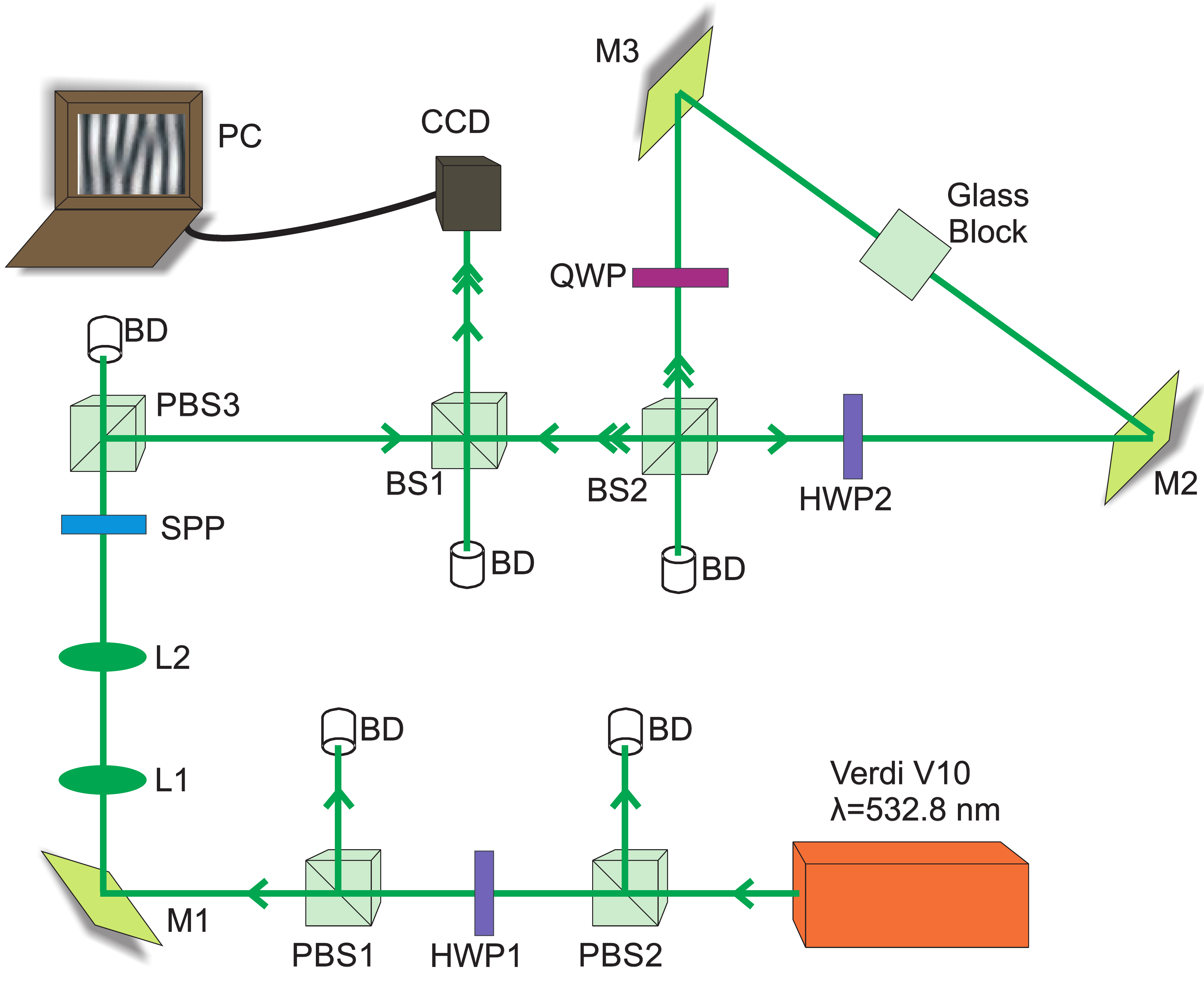}
  \caption{(Color online) Experimental setup for determination of TPCF. PBS: polarizing beam splitter; HWP: half wave plate; BD: beam dumpers; SPP: spiral phase plate; M: mirror; BS: beam splitter; QWP: quarter wave plate.}\label{fig:ExpSetupWDF}
\end{center}
\end{figure}

The experimental setup to find the TPCF is shown in Fig. \ref{fig:ExpSetupWDF}. For this study, we have used Coherent Verdi-V10 laser with wave-length $\lambda$=532.8 nm. The power of the laser is attenuated by rotating a half-wave plate (HWP1) placed between two polarizing beam splitters (PBS1 and PBS2). The remaining power was dumped on to beam-dump (BD). The lens combination of L1 and L2 (L2$<$L1) is used to reduce the beam-size, so that only $m=0$ modes are generated.
This beam is then passed through a spiral phase plate (SPP) of desired order to generate optical vortices. The vortex with the vertical polarization is coupled to the Shearing-Sagnac interferometer (SSI) that comprises the beam splitter BS2 and two mirrors, M2 and M3. A quarter-wave plate (QWP) and a half-wave plate (HWP2) are kept in common path for the quadrature selection. A glass block mounted upon a rotation stage is also kept in the common path to introduce shear in two transverse directions. Presence of beam splitter (BS1) ensures that both the clockwise (CW) and the counter-clockwise (CCW) fields experience one reflection from and one transmission through the beam splitter. This removes the effect of deviations from 50\% transmission. We have not used any neutral density filters as they were introducing noise in the interference fringes. The two counter-propagating beams are interfered and imaged using a CCD camera that is connected to computer PC. These interferograms without image processing are used to find the TPCF and the WDF.

\begin{figure}[h]
\begin{center}
  \includegraphics[width=2.5in]{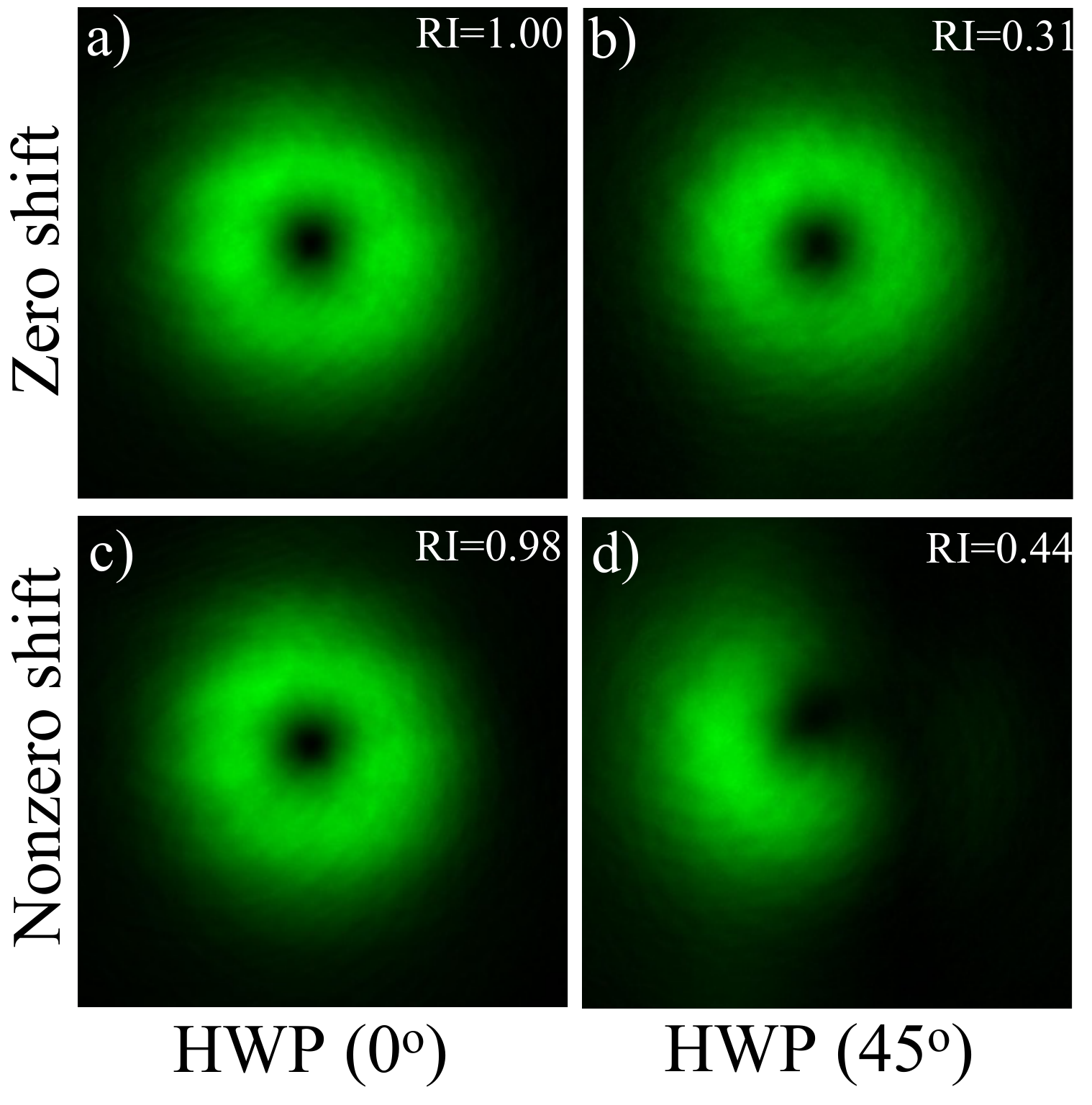}
  \caption{(Color online) Interferograms recorded with HWP at 0$^{\rm o}$ and 45$^{\rm o}$ for zero shear (a, b), for nonzero shear (c, d). The relative intensities (RI) of these interferograms are mentioned on the inset of each frame and are scaled for observation.} \label{fig:vortex}
\end{center}
\end{figure}

Before going to the shearing interferometry, we have calibrated the beam shear produced by the glass block for the Gaussian laser beam hosting the vortex. For detailed description, see appendix. The main part of our experiment is to determine the TPCF \cite{iaconis_direct_1996, kumar_information_2011}. For various tilts of the glass block, we have recorded the interferograms by keeping the fast axes of the QWP and the HWP parallel to the incident beams' polarization direction. In this orientation, the wave plates have no effect on the polarization of the optical beam, and both the CW and CCW propagating fields travel equal optical path lengths inside the SSI. The recorded interferograms contain the information of Re$[\Phi(X,\epsilon_x;Y,\epsilon_y)]$. Keeping the same lateral shear values, interferograms for Im$[\Phi(X,\epsilon_x;Y,\epsilon_y)]$ are taken after rotating the HWP by $\pi/4$ such that both the CW and CCW propagating fields rotate in polarization by 90$^\circ$. Similarly, the interferograms were obtained by tilting the glass cube to corresponding positions along $x$ and $y$ axes as suggested by Fig. \ref{fig:scaling}. The interferograms recorded for zero shear ($X=Y=0$) are shown in Fig. \ref{fig:vortex}(a, b) and for nonzero shear ($X=0.2$, $Y=0.0$) in Fig. \ref{fig:vortex}(c, d).

\section{Results and Discussion}\label{sec:result}
\begin{figure}[h]
\begin{center}
  \includegraphics[width=3.5in]{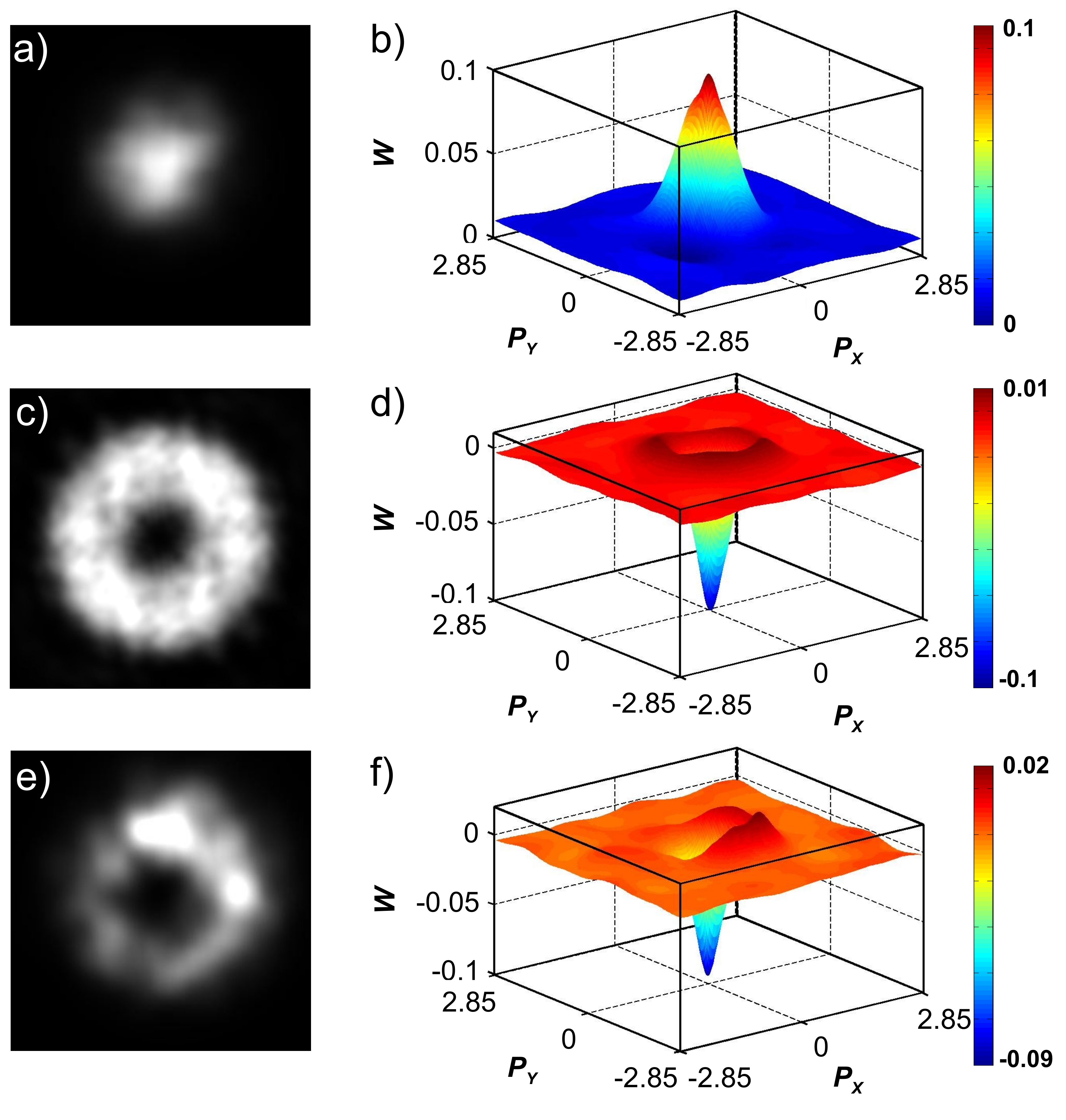}
  \caption{(Color online) Experimentally obtained the absolute value of TPCF (left column) and corresponding Wigner distribution function (right column) for Gaussian beam (first row) and optical vortex of topological charge $n=1$ at zero shear $X=Y=0$ (middle row) and nonzero shear $X=0.2$ and $Y=0$ (bottom row).}\label{fig:TPCF}
\end{center}
\end{figure}

The TPCF is obtained by subtracting the two individual beam components from the recorded interferograms. Figure \ref{fig:TPCF}(a) shows the TPCF of a Gaussian beam, while \ref{fig:TPCF}(c) and \ref{fig:TPCF}(e) show the same for optical vortex of topological charge $n=1$ at zero shear ($X=Y=0$) and nonzero shear ($X=0.2$, $Y=0.0$) respectively. To obtain the WDF, we have taken the Fourier transform of the experimentally obtained TPCF \cite{pratap_singh_wigner_2006}. Figure \ref{fig:TPCF}(b) shows the WDF of Gaussian beam while \ref{fig:TPCF}(d) and \ref{fig:TPCF}(f) show the WDFs for the optical vortex with topological charge $n=1$ at zero shear ($X=Y=0$) and nonzero shear ($X=0.2$, $Y=0.0$) respectively.

After obtaining the four WDFs at chosen shear values ($X1$, $Y1$), ($X2$, $Y1$), ($X1$, $Y2$) and ($X2$, $Y2$), the four dimensional addition was performed over $P_{X1}, P_{X2}, P_{Y1}, P_{Y2}$ axes to determine $B$ as defined in Eq. \ref{eq:eightdimint}. The experimentally obtained WDF is a two-dimensional ($P_X$, $P_Y$) function, keeping two dimensions ($X$, $Y$) to be constant. However, after addition of four WDFs, the Bell parameter $B$ becomes a four-dimensional function ($P_{X1}$, $P_{X2}$, $P_{Y1}$, $P_{Y2}$) with other four dimensions ($X1$, $X2$, $Y1$, $Y2$) being fixed. Equation \ref{eq:eightdimint} shows the generation of a four-dimensional matrix after adding the four two-dimensional functions. However, proper axes should be considered while adding. The maximum value of $B$ was determined to verify the violation of Bell's inequality.

\begin{figure}[h]
\begin{center}
  \includegraphics[width=3.1in]{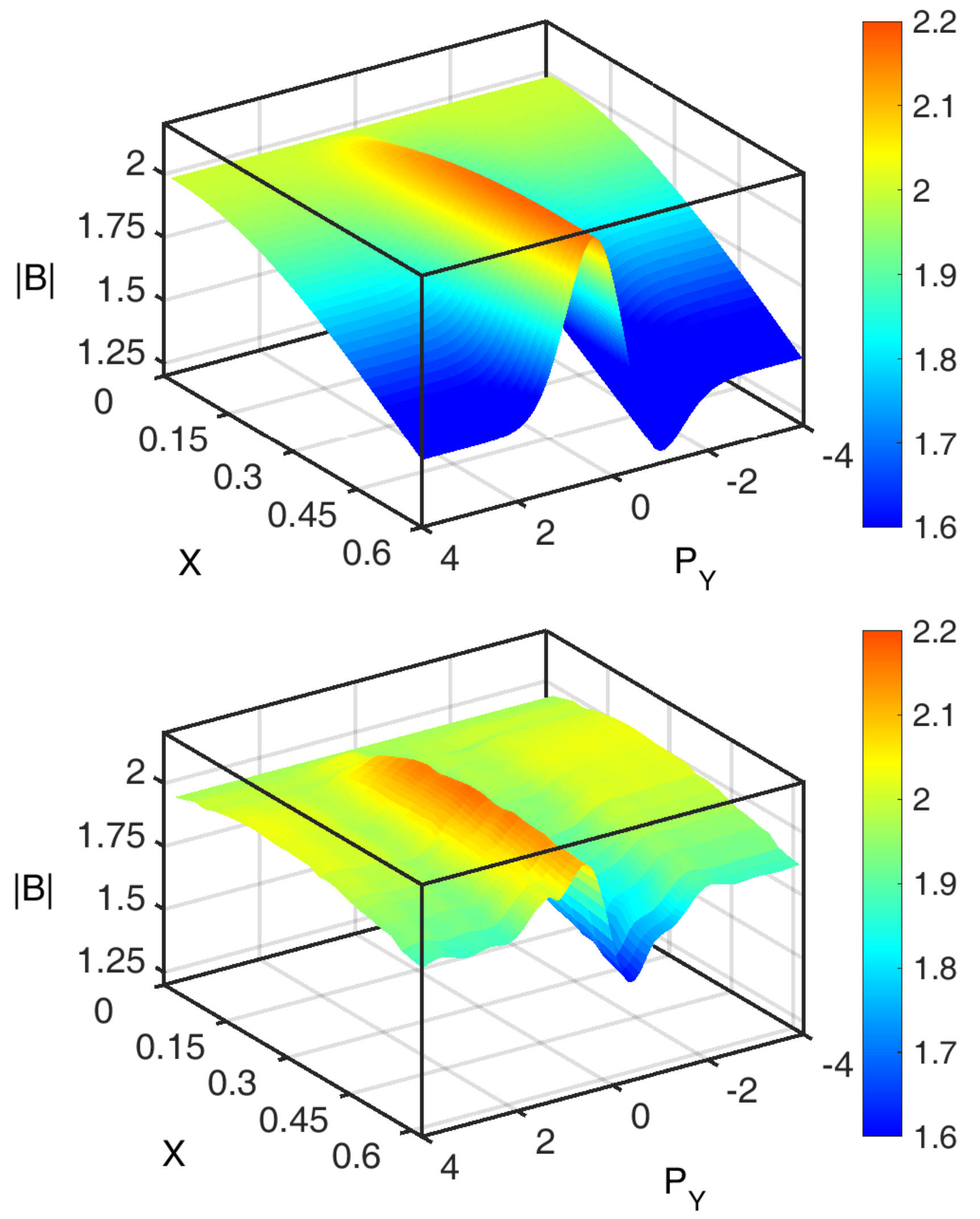}
  \caption{(Color online) Variation of $|B|$ with $X$ and $P_Y$ (Eq. \ref{eq:order1}) for $n=1$. Theoretical (top) and experimental (bottom) for $X1$=0, $P_{X1}$=0, $X2$=$X$, $P_{X2}$=0, $Y1$=0, $P_{Y1}$=0, $Y2$=0 and $P_{Y2}=P_Y$.}\label{fig:plot}
\end{center}
\end{figure}

Considering $X1$=0, $P_{X1}$=0, $X2$=$X$, $P_{X2}$=0, $Y1$=0, $P_{Y1}$=0, $Y2$=0, $P_{Y2}=P_Y$, the 2D surface plot of $|B|$ varying with $X$ and $P_Y$, described by Eq. \ref{eq:order1}, is shown in Fig. \ref{fig:plot}. From the plot, location of the maximum of $|B|$ has been determined that matches with the theory. The $|B_{max}|$ obtained from Fig. \ref{fig:plot} is 2.1649$\pm$0.0079, which indicates that the continuous variables of optical vortex field are non-separable.

Due to the common path interferometry, the multiple back reflections from various optical components reaching the camera are one of the main issues faced during the experiment. The small fluctuations in Fig. \ref{fig:plot} can be attributed to the finite number of data points while recording the TPCF experimentally and discrete Fast Fourier transform of that data while determining the WDF.

Corresponding to vortices of order $n$=1, 2 and 3 as well as for the Gaussian beam ($n$=0), Eq. \ref{eq:eightdimint} has been solved numerically to obtain the parameter values for maximum $B$. The obtained $X1$, $X2$, $Y1$, $Y2$ values have been used to select the desired shear for the experimental measurements of the TPCF, which provides us the WDF through Fourier transform. The obtained shear values for different orders are listed in Table \ref{table}.

\begin{table}[h]
\begin{tabular}{|p{0.8cm}|p{1.2cm}|l|}
\hline 
$n$ & $|B_{max}|$ & ($X1, X2, Y1, Y2$) \\ 
\hline 
0 & 2.00 & (0.00, 0.58, 0.00, 0.00) \\ 
\hline 
1 & 2.24 & ($-$0.07, 0.40, $-$0.05, 0.26) \\ 
\hline 
2 & 2.35 & (0.09, $-$0.40, 0.00, 0.00) \\ 
\hline 
3 & 2.40 & ($-$0.09, 0.35, $-$0.01, 0.06) \\ 
\hline 
\end{tabular}
\caption{Theoretical shear values providing $|B_{max}|$.} \label{table}
\end{table}

\begin{figure}[h]
\begin{center}
  \includegraphics[width=3.1in]{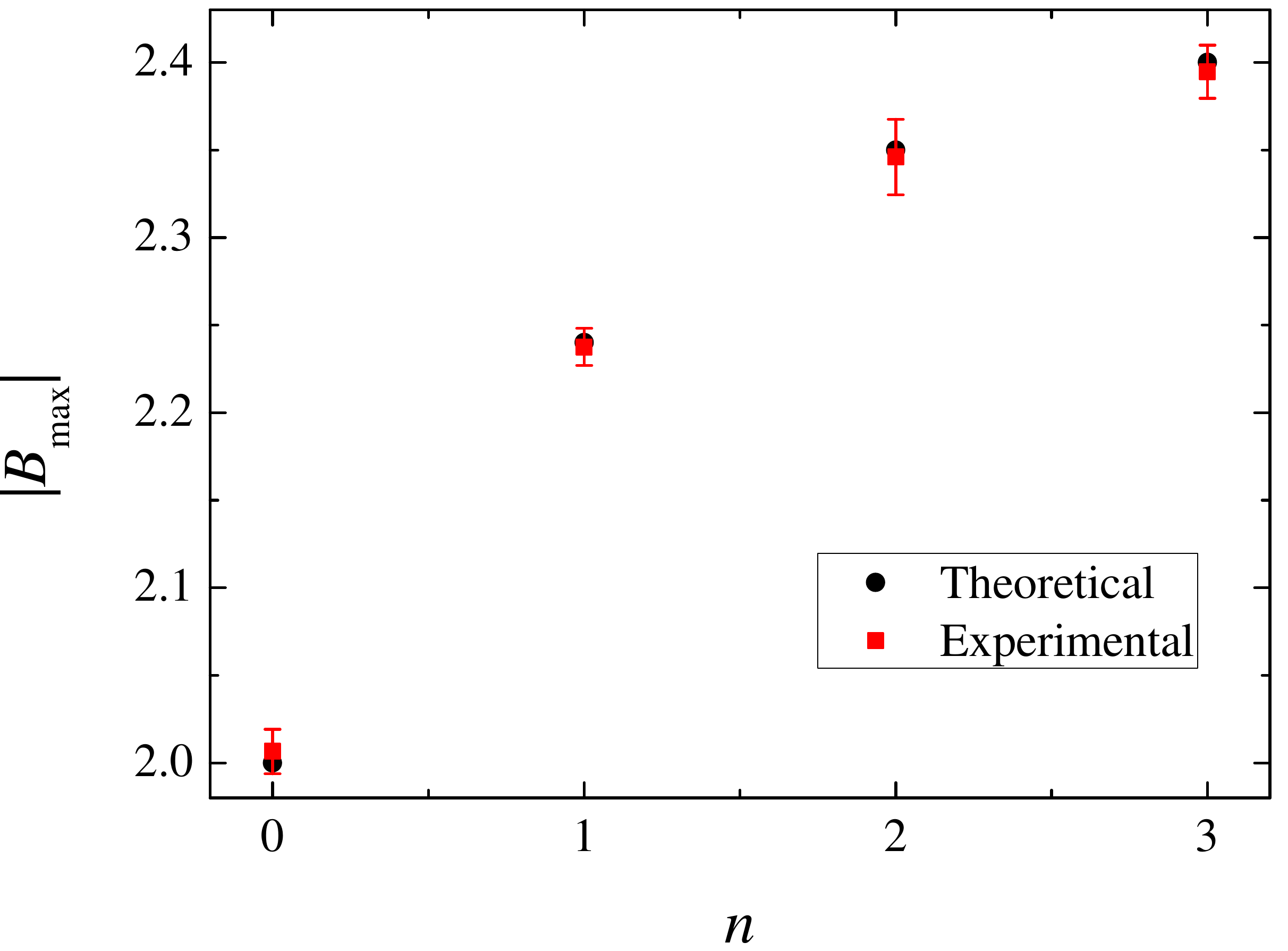}
  \caption{(Color online) Variation of $|B_{max}|$ with the order of vortex ($n$) obtained at parameters mentioned in Table \ref{table}.}\label{fig:BvsN}
\end{center}
\end{figure}

Figure \ref{fig:BvsN} shows the variation of Bell's inequality violation ($|B_{max}|$) for a Gaussian beam and the optical vortices of order $n=$1--3. From Fig. \ref{fig:BvsN}, it is clear that there is no Bell's inequality violation for the Gaussian beam. However, for the optical vortex beams, the Bell's inequality has been violated. The amount of Bell's violation increases with the increase in order of the vortices. The amount of entanglement increases with the order of an optical vortex due to the increase in Bell's violation parameter ($B_{max}$). Since the earlier results also point to an increase in information entropy \cite{kumar_information_2011} - a measure of entanglement - with the order of vortex, therefore, in the present case, $|B_{max}|$ can be used to obtain the degree of entanglement. To verify our numerical solution for parameters giving maximum $B$, we have also performed experiments around the point of $|B_{max}|$ and observed that the amount of Bell's violation decreases as we move away from the point of maxima. It verifies our numerical solution for parameters maximizing $B$.

To estimate the experimental error, twenty five sets of interferograms were recorded. In every set of experiment, four WDFs were determined and for each WDF, four sets of inteferograms corresponding to two individual beams and real as well as imaginary components of the TPCF were recorded. $|B_{max}|$ was calculated for each set of experiments. The $|B_{max}|$ used in Fig. \ref{fig:BvsN} is the average of twenty five $|B_{max}|$ determined from each set of experimental interferograms. Error bars are the standard deviations around mean for twenty five values of $|B_{max}|$.

\section{Conclusion}\label{sec:conclusion}
In conclusion, we have experimentally verified the quantum like classical entanglement for optical vortex beams. We have shown that these classical beams violate Bell's inequality for continuous variables of position and momentum. The extent of violation of Bell's inequality increases with the increase in topological charge. The violation of Bell's inequality in phase-space ($x$, $p_x$; $y$, $p_y$) clearly shows the existence of spatial correlation properties similar to entanglement in quantum systems for optical vortices which is different compared to the Gaussian beam. One must be able to see this type of entanglement for electron vortex beams also due to the generic nature of vorticity.

\section{Acknowledgments}
Authors wish to acknowledge Prof. G. S. Agarwal for useful discussion and to Prof. J. M. Raimond for his critical comments.

\section*{APPENDIX: Calibration of glass cube in shearing Sagnac interferometer}
In this appendix, we are presenting the method for calibrating the shearing Sagnac interferometer, which has been used to study the Bell's inequality violation expressed in terms of the WDF for optical vortices.

\begin{figure}[h]
\begin{center}
  \includegraphics[width=2in]{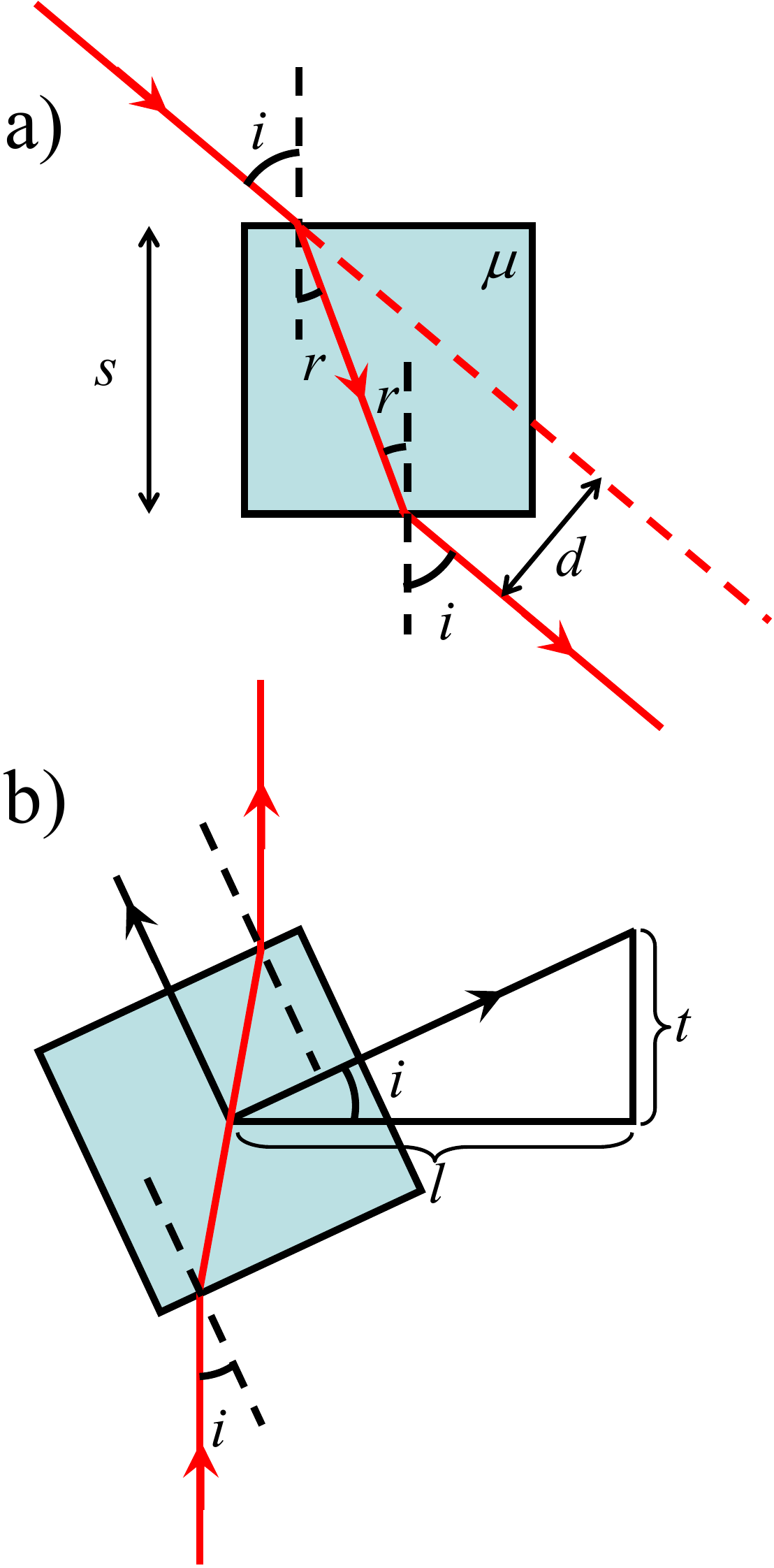}
  \caption{(Color online) The ray diagrams of laser beam passing through the cube.}\label{fig:drawing}
\end{center}
\end{figure}

Before describing the experiment, we have first modeled the shift in beam due to the glass block by using Snell's law of refraction. To obtain the calibration curve, we have considered the laser beam falling at an incident angle $i$ on the cube. The ray diagram representing the cube and the beam shift is shown in Fig. \ref{fig:drawing}(a) while the mounting arrangement of the cube is shown in Fig. \ref{fig:drawing}(b). In the present case, the separation between the center of the cube and the linear scale on the mount is $l=5.2$ cm. $t$ is the linear scale on the mount.

Considering the refractive index of cube as $\mu$ and its thickness as $s$, from the Snell's law of refraction, we can write

\begin{equation}
\mu = \frac{\sin(i)}{\sin(r)}
\end{equation}
\begin{equation}\label{eq2}
d=s\times\frac{\sin(i-r)}{\cos(r)}
\end{equation}
From Fig. \ref{fig:drawing}(b), one can write
\begin{equation}\label{eq3}
\tan(i)=\frac{t}{l}
\end{equation}
when $i\rightarrow 0$, $r\rightarrow 0$ and using Eq. \ref{eq3}, the shift in the beam can be written as
\begin{equation}\label{final}
d=\frac{st}{l}\left( 1-\frac{1}{\mu} \right)
\end{equation}
For smaller angle of incidence i.e. within the paraxial approximation, the shift induced by the cube varies linearly with the incidence angle $i$. In the present experiment, maximum tilt given to the cube was $\sim3^{\rm o}$, that corresponds to the linear variation in the shift.

The optical vortex generated through spiral phase plate (SPP) and interference pattern of vortex with itself is given in Fig. \ref{fig:vortex}. The images shown are unprocessed images. One can notice the clarity of the vortex and the interference fringes.
\begin{figure}[h]
\begin{center}
  \includegraphics[width=3.1in]{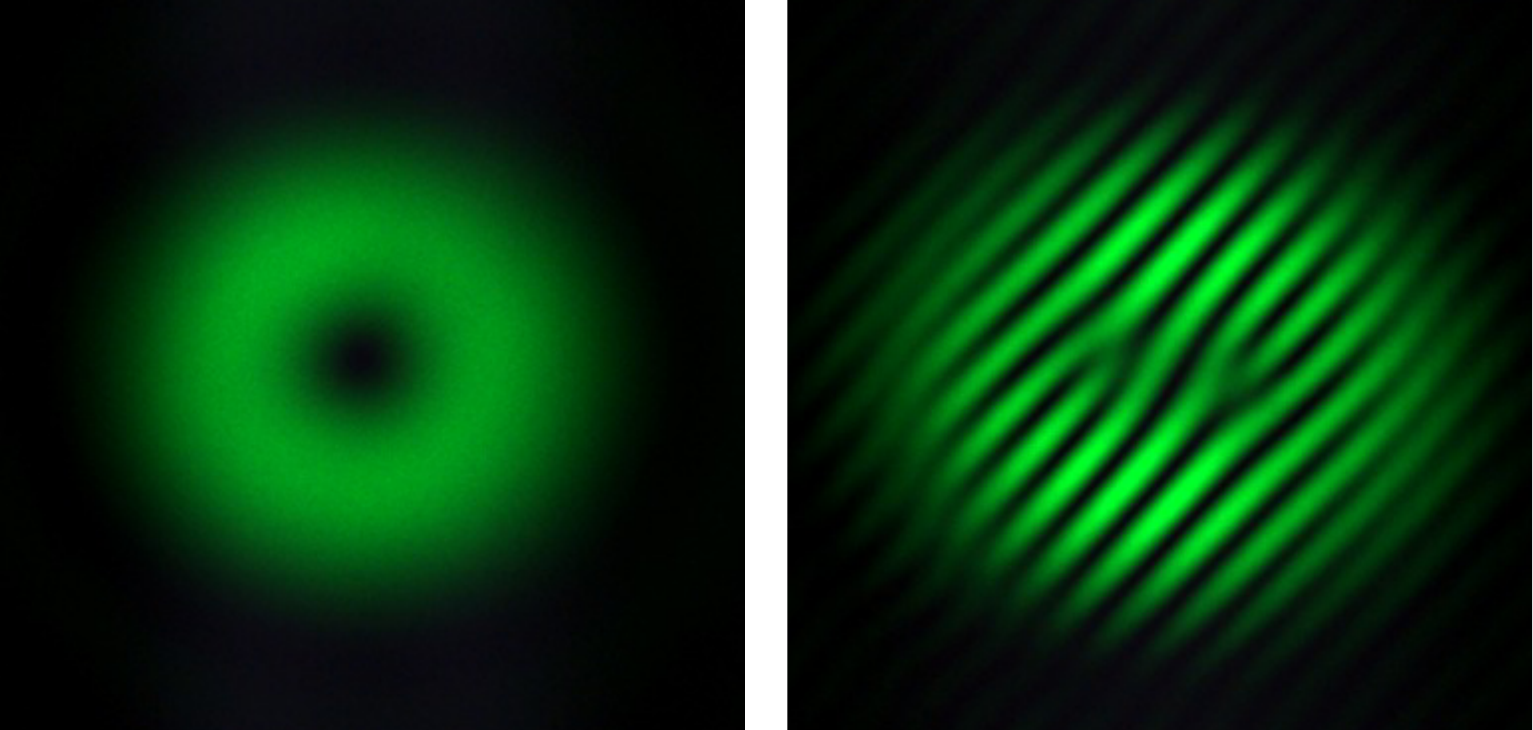} 
  \caption{(Color online) Optical vortex (left) and its inteference with itself (right).}\label{fig:vortex}
\end{center}
\end{figure}

For calibrating the shift due to the glass cube in shearing Sagnac interferometer (SSI) (Fig. \ref{fig:ExpSetupWDF}), we put one polarizer inside the SSI, near to the cube. The CW beam passes through QWP, polarizer and then HWP while the CCW beam passes through HWP, polarizer and then QWP. As the vortex beam entering the SSI is vertically polarized and the HWP is kept at ($\pi/4$), the CCD records only CW and CCW beams at the angle of polarizer at 0 or $\pi/2$ respectively. The presence of QWP at 0$^{\rm o}$ will not have any effect on the beam selection. In this way, the two beams were selected by the polarizer.

\begin{figure}[h]
\begin{center}
  \includegraphics[width=3.1in]{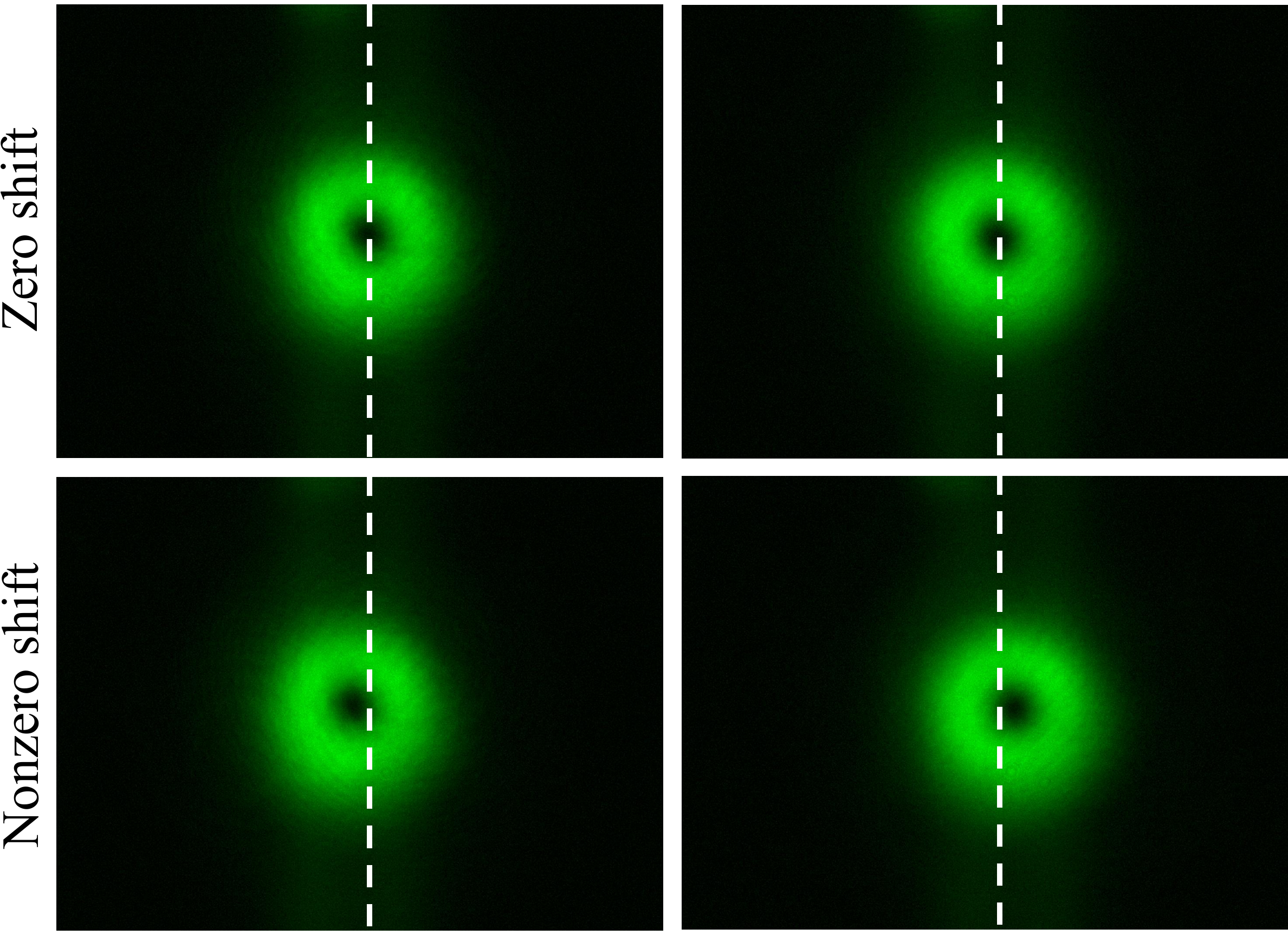} 
  \caption{(Color online) Left and right columns correspond to the CW and CCW beams respectively. The vertical dashed lines show the shift in two beams due to the shear provided by tilting the glass block.}\label{fig:beamshift}
\end{center}
\end{figure}

\begin{figure}[h]
\begin{center}
  \includegraphics[width=3.1in]{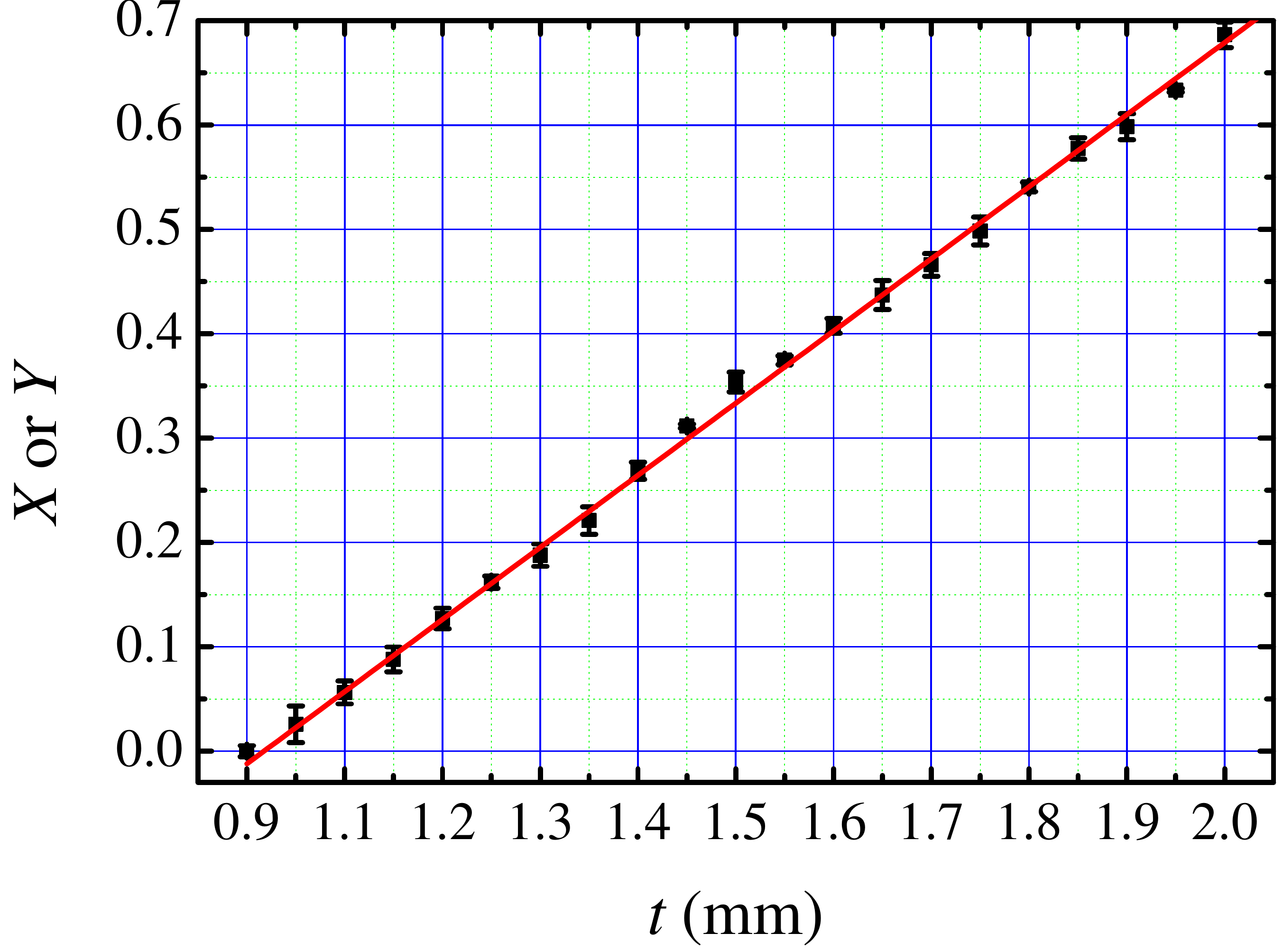} 
  \caption{(Color online) Calibration curve for dimensionless shear ($X$, $Y$) in the SSI. The $x$-axis ($t$) denotes the position on linear scale of the mount on which glass cube was mounted. The red line is a linear fit to our experimental data.}\label{fig:scaling}
\end{center}
\end{figure}

Starting with a zero shear, we provide gradual shear to the beam with the glass block mounted on a rotation stage with a linear scale. The tilt of cube is varied in equal steps. To determine the shear between two beams, we recorded the intensities of two beams in a CCD camera with pixel size 4.65 $\mu$m. Figure \ref{fig:beamshift} shows the CCD frames recorded at zero as well as nonzero shear. One can observe the shift in position of the beams in the image frames. The width $w$ of the vortex falling on the CCD was determined using the 2D curve-fitting.

The scaled shear was determined using scaling relation Eq. \ref{eq:scaled_shift} which correspond to the linear scale on rotation stage of the glass block. The error bar denotes the standard deviation of twenty five measurements. The calibration curve matches with the linear fitting, as derived in Eq. \ref{final}. The amount of shear as a function of linear scale of the rotation stage has been shown in Fig. \ref{fig:scaling}. This calibration curve is further used in the experiment for the Bell's inequality violation in terms of the WDF for optical vortex beams.


\end{document}